\documentclass[12pt]{article}
\usepackage[dvips]{epsfig}
\usepackage{cite}

\title{Field-induced interaction of a pseudoscalar particle with
 photon in a magnetized $e^-e^+$ plasma}

\author{N.V.~Mikheev\thanks{E-mail address:mikheev@univ.ac.ru},
E.N.~Narynskaya\thanks{E-mail address:elenan@univ.ac.ru}\\
{\small\it Division of Theoretical Physics, Department of Physics,}\\
{\small\it Yaroslavl State University, Sovietskaya 14,}\\
{\small\it 150000 Yaroslavl, Russian Federation.}}

\date{}

\begin{document}

\maketitle

\begin{abstract}

The effective interaction of a pseudoscalar particle with  photon
in plasma with the presence of a constant uniform magnetic field
 is investigated. It is shown that under some physical conditions
the effective coupling between pseudoscalar particle and photon
does not  depend on medium  parameters and particles momentum.The
probability of the familon decay into photon pair in a
strongly magnetized degenerate ultrarelativistic plasma is
calculated.

\end{abstract}

\def\beq{\begin{equation}}
\def\eeq{\end{equation}}
\def\bd{\begin{displaymath}}
\def\ed{\end{displaymath}}
\def\f{\phi \to  \gamma}
\def\g{g_{\phi\gamma}}

\section{Introduction.}

\indent\indent For a long time there has been a stable interest to
investigations of quantum processes in an external active medium,
which can be presented by plasma  as well as a magnetic field.
Actually, both components  of the active medium could exist in
astrophysical objects. By this means, the interest to studies of
elementary  particle physics under extreme conditions, dense
plasma, strong magnetic field and/or high temperature, is caused,
  in particular,  by  possible astrophysical applications. Of special
interest  are the  processes with participation of a light weakly
interacting particle. It should be noted that in previous studies
of quantum processes in a dense stellar matter the main attention
was given to the neutrino processes. This is due to the fact that
neutrino plays determining role in astrophysical cataclysms like a
supernova  explosion and a coalescence of neutron stars.

However, the investigations of the other weakly
interacting particles physics (axion, familon etc.)  could  be of
interest for astrophysical applications also~\cite{Raffelt1}. In
view of weakly interaction with matter such particles could give a
perceptible effect on the dynamics of a cooling star. In
particular,
 the emission processes of  these
particles could give an additional contribution into the star
energy losses~\cite{Borisov,Kacherliess,Vassilevskaya,Mikheev1}.
The other interesting effect from astrophysics viewpoint is the
possible asymmetry of emission of the weakly interacting particles
from a supernova caused by the presence of an external magnetic
field. This asymmetry could lead to the reactive force and, as a
result, to the initial impetus of a pulsar
(kick-velocity)~\cite{Mikheev1}.

In this paper we study the interaction of a pseudoscalar particle
with  photon in the electron-positron plasma with the presence of
an external magnetic field. As a pseudoscalar particle we will
consider the familon, the Nambu-Goldstoun boson arising as a
result of the spontaneous breakdown  of a  global symmetry between
fermion generations~\cite{Wilczek,Anselm}. Notice, that
interaction between familon and photon becomes possible  in  an
external magnetic field only because of the fact that familon does
not have both anomalous $\Phi G \tilde G$ and
 $\Phi F \tilde F$ coupling in vacuum ($G$ and $F$ are gluonic  and electromagnetic
fields, respectively).

The field-induced effective  familon-photon interaction is
described by the loop Feynman diagram shown in Fig.1. and  can be
presented as
\beq
  L_{\phi \gamma}= g_{\phi\gamma} \,
 \, \tilde F^{\alpha\beta} \, (\partial_\beta \, A_\alpha) \, \Phi.
\label{eq:lag1} \eeq
Here $A_\mu$ is the four-potential of the quantized
electromagnetic field, $\Phi$ is the familon field, $\tilde
F^{\alpha\beta}=\frac{1}{2}
\varepsilon^{\alpha\beta\sigma\tau}F_{\sigma\tau}$ is the dual
tensor of the external magnetic field $F_{\sigma\tau}$,
$g_{\phi\gamma}$ is the effective familon-photon coupling in a
magnetized plasma.

The Lagrangian (\ref{eq:lag1}) leads to the amplitude of the
familon $\to$  photon transition in the following form
\beq
 M_{\f} = i g_{\phi\gamma} (\varepsilon^* \tilde F q),
\label{eq:amp1} \eeq
where $q_\mu = (\omega, {\vec q})$ is the photon 4-momentum,
$\varepsilon_\mu$ is the photon polarization 4-vector.

In a magnetized plasma, along with the loop-channel of the familon
$\to$ photon  transition the additional channel caused by  the
 presence of plasma, becomes possible, namely, the Compton-like
familon-photon "forward scattering" on plasma electrons and
positrons (Fig.2). Notice, that the contribution of this process
to the effective interaction between pseudoscalar particle and
photon  was not taken into account
previously~\cite{Vassilevskaya}. However, as it will be shown
below, under some physical conditions
 the plasma contribution
into the effective coupling  $\g$ could appear to be much more
than the field one.

So, in a magnetized plasma the amplitude of the transition $\f$
can be written in general case as the sum of the field and plasma
contributions
\beq
 M_{\f} = M^{F} + M^{Pl}. \label{eq:amp}
\eeq
\begin{figure}[t]
%
%\centerline{\includegraphics{fig1.eps}} 
\centerline{\psfig{file=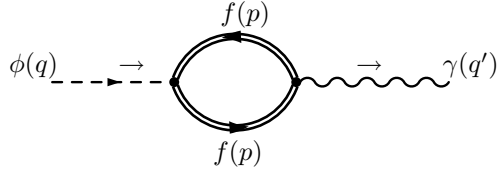}} 
\caption{The field-induced
interaction of a pseudoscalar particle and photon. Double lines
indicate, that the influence of an external magnetic field is
taken into account in the propagator of virtual fermions.}
\label{fig:fig1}
\end{figure}

\begin{figure}[b]
%
%\centerline{\includegraphics{fig2.eps}}
\centerline{\psfig{file=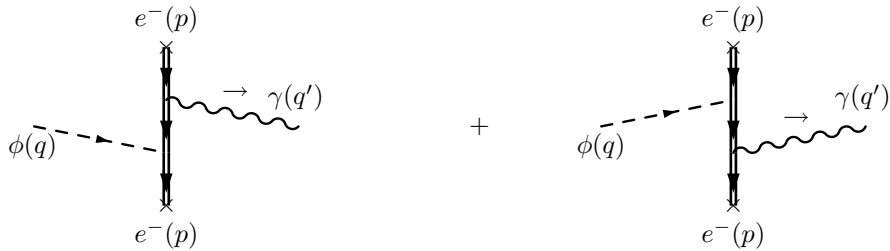}}
\caption{The diagrams
describing the Compton-like  familon $\to $
 photon "forward  scattering" on  plasma electrons.}
\label{fig:fig2}
\end{figure}

Further we will calculate the plasma and field contributions into
the  effective  familon-photon coupling in plasma with the
presence of  a magnetic field.

\section{ Field-induced part of the effective coupling
between pseudoscalar particle and photon in a magnetized plasma.}

\indent\indent The field-induced part of the effective coupling
$\g$ can be derived from the diagram shown on Fig.1, where the sum
over all virtual fermions is considered. The main contribution
into this sum comes from an electron as the particle with the
maximal specific charge $e/m_e$, which is the most sensitive to
the external magnetic field influence.

The Lagrangian describing the familon-electron interaction is
\beq  L =  \frac{c_e}{{\it v}_\phi} \,\,(\bar \Psi_e \gamma^\mu \,
\gamma_5 \Psi_e) \,\, \partial_\mu \Phi,
  \label{eq:lag-2} \eeq
where $c_e$ is a model dependent parameter of the order of unit,
 ${\it v}_\phi$ is the horizontal symmetry breaking scale,
 $\Psi_e$ is the electron field.

In the second order of the perturbation theory, the $S$-matrix
element corresponding to the  diagram shown in Fig.1 can be
written as
\beq S^F  =  \frac{-iec_e}{2{\it v}_\phi V \sqrt{\omega \omega'}}
\,\, q_\mu\,\varepsilon_\nu \int d^4x \, d^4 y \, Tr \{S(y,x) \,
\gamma^\mu \, \gamma_5 \, S(x,y)  \, \gamma^\nu \} \, e^{i(q'y)}
\, e^{-i(qx)},
 \eeq
where $q^\alpha=(\omega, {\vec k})$ and  $q'^\alpha=(\omega', \vec
k')$  are the 4-momenta of  the familon and photon respectively,
$S(x,y)$ is the propagator of virtual electron in the loop, $e>0$
is the  elementary charge.

The electron propagator  in the constant  uniform magnetic field
  can be presented in the form~\cite{Izicson}
\bd S (x, y)  =  e^{i \Phi (x, y)} \, S (z), \ed
here $z_\mu = x_\mu - y_\mu$.

The translational and gauge non-invariant part of the propagator
is separated in the phase $\Phi(x,y)$ which can be defined in
terms of an integral along an arbitrary contour
 \beq
\Phi (x, y)  =   e \, \int_y^x \! d\xi^\mu \,
            \left [ A_\mu (\xi) +
           \frac{1}{2} F_{\mu \nu} (\xi - y)^\nu \right ].
           \label{eq:phase}\eeq

Taking into account that in the case of two-vertex fermion loop,
the sum of the phases arising  from two fermion propagators is
zero
\bd \Phi (x, y) + \Phi (y, x) = 0, \ed
one can define the amplitude of the process by the standard
manner~\cite{Ber}
\beq S = \frac{i (2 \pi)^4 \delta^4 (q - q')}{2 \omega V} \, M,
\label{eq:amp-def} \eeq
where the field-induced part of the amplitude can be presented in
the form
\beq M^F = \frac{-ec_e}{{\it v}_\phi} \,\,q_\mu \,
\varepsilon^*_\nu \,\, \int d^4 z \, e^{-i(qz)} \,
 \, Tr \{ S(-z) \, \gamma^\mu  \, \gamma_5 \, S(z)\, \gamma^\nu \}.
\label{eq:Mfield-1} \eeq

 The one-loop two-point amplitude in an external magnetic
field was intensively investigated
previously~\cite{Daas,Raffelt2,Kuznetsov}. The expression for the
amplitude of the familon-photon conversion could be extracted, for
example, from the book~\cite{Kuznetsov}, where in the section 4
the field-induced one-loop amplitude  of the transition
 $j \to f \bar f \to j'$  for various
combinations of scalar, pseudoscalar, vector and pseudovector
interactions of general currents $j$ and $j'$ with fermions are
presented. In particular, from  the formula (4.22) of
\cite{Kuznetsov} corresponding to the axial-vector and vector
vertices we obtain
\begin{eqnarray}
 M^{F} &  =  &  \frac{2 i  \alpha  c_e }{ \pi {\it v }_\phi}\,
 ( \varepsilon^*  \tilde F q)\, \,\int\limits_0^1 du \, \int\limits_0^\infty d \tau \, e^{ -
\Omega(u,\tau)},
\label{eq:Mfield-2}\\
\Omega(u,\tau) & = & \tau \left ( 1 -\frac{q^2_\parallel}{m^2_e}
\,{1 - u^2 \over 4} \right ) - \frac{q^2_\perp}{2 m^2_e} \left (
\frac{ch(\eta u \tau) - ch(\eta  \tau)}{\eta sh(\eta \tau)} \right
), \nonumber
\end{eqnarray}
where  $ \varphi_{\alpha\beta}$ is the dimensionless external field
tensor, $\varphi_{\alpha\beta} = F_{\alpha\beta} / B$, $\tilde
\varphi_{\alpha\beta} = \frac{1}{2}
 \varepsilon_{\alpha\beta\rho\sigma} \varphi^{\rho\sigma}$ is  the dual tensor,
$\beta = eB$, $\eta = \beta/m^2_e=B/B_e$ ($ B_e = m_e^2/e \simeq
4.41\times10^{13} $ G)~\footnote{ We use natural units in which
$c=\hbar=1$.},
 $q^2_\parallel = (q \tilde \varphi \tilde \varphi q) = q^2_0 -
 q^2_3$,
 $q^2_\perp = (q  \varphi \varphi q)  = q^2_1 + q^2_2$,
$q^2 = q^2_\parallel - q^2_\perp$ (it is assumed that a magnetic
field is directed
 along the third axis, $\vec B = (0,0, B$)),
 $m_e$ is the  electron mass, $\alpha = e^2/4\pi$.

Comparing the expression (\ref{eq:Mfield-2}) with the amplitude
(\ref{eq:amp1}) we find the field contribution into the effective
coupling of the familon-photon interaction in the form
\beq
 g^{F}_{\phi\gamma}  =   \frac{ 2 c_e  \alpha }{ \pi {\it v }_\phi}\,
\,\int\limits_0^1 d u \, \int\limits_0^\infty d \tau \, e^{ -
\Omega(u,\tau)}. \label{eq:gfield1} \eeq

It should be emphasized that, strictly speaking, the field-induced
amplitude of the processes featuring pseudoscalar particles and
therefore the effective coupling (\ref{eq:gfield1}) are a
non-uniqueness  physical quantity because of the Adler triangle
anomaly. Since the familon-photon interaction is free from the
Adler anomaly, the effective familon-photon coupling must
disappear in the local limit when $q^2=0$. Therefore obtaining the
physically  correct result it is necessary to perform the
procedure of the substraction of the Adler anomaly which is
reduced to the substraction from the field-induced coupling
(\ref{eq:gfield1}) the one in the limit $m_e \to \infty$
\beq
 g^{F}_{\phi\gamma}  =   \frac{2 c_e  \alpha }{ \pi {\it v }_\phi}\,
\left ( \,\int\limits_0^1 d u \, \int\limits_0^\infty d \tau \, e^{ -
\Omega(u,\tau)} - 1 \right ). \label{eq:gfield2} \eeq

The expression (\ref{eq:gfield2}) describes the field induced part
of the effective coupling.

\section{ Plasma part of the effective coupling  between
pseudoscalar particle and photon in a magnetized plasma.}

\indent\indent The plasma contribution into the effective
familon-photon coupling is caused by the Compton-like
familon-photon "forward scattering" on plasma electrons and
positrons (Fig.2). With  the  Lagrangian  (\ref{eq:lag-2}) we find
the S-matrix element corresponding to
 the diagrams of Fig.2 in the form
\begin{eqnarray}
S_{e^-}^{Pl}  & = &  \frac{i e c_e }{2 {\it v }_\phi V\sqrt{\omega
\omega'}} \, q_\mu \, \varepsilon^*_\nu \,
 \sum_{n=0}^\infty \sum_s \int
                    d^4 x \,\, d^4 y \,\, dn_{e^-}\,e^{-iqx}\, e^{iq'y}
\label{eq:Splasma-1} \times\\[2mm]
    & \times  &
 Tr (  \bar \psi_e(p,x)  \,  \gamma^\mu \, \gamma_5 \,S(x,y)\,
   \gamma^\nu \,  \psi_e(p,y)\, + \, \bar \psi_e(p,y)  \,  \gamma^\nu \,S(y,x)\, \gamma^\mu \,\gamma_5
    \psi_e(p,x)) ,
     \nonumber
\end{eqnarray}

where $q^\alpha=(\omega, {\vec k})$ and  $q'^\alpha=(\omega', \vec
k')$  are the 4-momenta of  the familon and photon respectively,
  $p^\mu=(E_n, \vec p)$ is the electron 4-momentum, $E_n = \sqrt{p_z^2 +
2n\beta + m^2_e}$ is the energy, $n$ is the Landau level number,
$\psi_e$ is the solution of the Dirac equation in magnetic field,
$S(x,y)$ is the electron propagator, $dn_{e^-}$  is the phase
space element of plasma electrons. In the external magnetic field,
the number of plasma electrons  in the gauge $A^\mu=(0,0,Bx,0)$ is
defined as
\bd dn_{e^-} = \frac{dp_2 dp_3}{(2\pi)^2}\,L_2 \, L_3\,f(E_n, \mu). \ed
Here $L_1,L_2$ are the auxiliary parameters which determine
 the normalization volume $V=L_1L_2L_3$, $ f(\ E_n, \mu) $ is the
electron distribution function, which has the following form in the plasma rest frame
\bd f(E_n, \mu)= \frac{1}{e^{(E_n - \mu)/T} + 1}, \ed
where $\mu $ and $T$ are the plasma chemical potential and
temperature respectively.

The S-matrix element of  the familon-photon "forward scattering"
on plasma positrons can be obtained from (\ref{eq:Splasma-1}) by
substitution $p \to -p$  in the solution of the Dirac equation and
$\mu \to -\mu$ in the distribution function.

The electron wave function in the above-mentioned  gauge can be
written as~\cite{Axiezer}
\beq \psi_e(p,x)   =  \frac{u_s(p,\xi) \,  e^{-i(E_n x_0 - p_2 x_2 - p_3 x_3)}}{\sqrt{2E_n(E_n + m_e)L_2
L_3}}, \label{eq:vf} \eeq
where the bispinor amplitude, corresponding to the two projections
of the electron spin on the field direction, $s = \pm1$, are given
by
\bd
u_{s=-1}(p,\xi)=
\left(
\begin{array}{c}
0 \\
(E_n + m_e)\,\,V_{n} (\xi) \\
-i \sqrt{2\beta n} \,\,V_{n-1} (\xi) \\
-p_3 \,\, V_n (\xi)
\end{array}\right ), \,\,\,
u_{s=+1}(p,\xi)=
\left(
\begin{array}{c}
(E_n + m_e)\,\,V_{n-1} (\xi) \\
0\\
p_3 \,\, V_{n-1} (\xi) \\
i \sqrt{2\beta n} \,\,V_{n} (\xi) \\
\end{array}\right ). \nonumber
\ed
Here $ V_n(\xi) =  \frac{\beta^{1/4}}{\sqrt{2^n n! \sqrt \pi}}  e^{-\xi^2/2}
H_n(\xi)$ , $H_n(\xi)$ is the Hermite polynomials,
 $\xi$ is the dimensionless variable,
$\xi  =  \sqrt{\beta} \left( x_1 + p_2/\beta \right)$.

The translational and gauge non-invariant phase  of the electron
propagator (\ref{eq:phase}) in the considered  gauge is reduced to
the simple form
\bd
\Phi (x, y) = -\frac{\beta}{2} (x_1 + y_1) (x_2 - y_2).
\ed

The translational and gauge invariant part of the propagator
  $S(z)$  has several representations.  For our purposes it is
  convenient  to take it in the form
\begin{eqnarray}
 S(z) &=& - \frac{i}{4 \pi} \int \limits^{\infty}_{0}
\frac{d\tau}{th \tau}\, \int \frac{d^2 p'_\parallel}{(2 \pi)^2}
\Biggl \{ [ (p'\gamma)_\parallel + m_e]\,\Pi_{-}\,(1 + th \tau) +
\nonumber\\
&+& [(p' \gamma)_\parallel  + m_e]\,\Pi_{+}\,(1 - th \tau) - \frac{i
\beta \, (z \gamma)_\perp}{2\, th \tau}\, (1 - th^2 \tau) \Biggr \}
\nonumber \\
& \times & \exp\left(- \frac{\beta z_{\perp}^2}{4\, th \tau} -
\frac{\tau(m^2_e - p_{\parallel}'^2)}{\beta} -
i(p'z)_\parallel\right), \label{eq:prop1}
\end{eqnarray}
where  $z_\mu = x_\mu - y_\mu$, $p'_\mu$ is the four-momentum of
the virtual electron, $d^2p'_{\parallel}=dp_0'dp'_3$,
 $\Pi_{\pm} = \frac{1}{2} (1  \pm
i \gamma_1\gamma_2)$, $ (p'\gamma)_\parallel = (p'_\mu
\gamma^\mu)_\parallel =  p'_0\gamma_0 - p'_3\gamma_3$, $ (z
\gamma)_\perp = (z_\mu \gamma^\mu)_\perp = z_1\gamma_1 +
z_2\gamma_2$.

After performing the partial integration  of the Eq.(\ref{eq:Splasma-1})
 over 4-coordinates  $x$ and $y$, impulses of
virtual electron and the  second component of plasma electron's
momentum $p_2$, the $S$-matrix element of the familon-photon
transition with coherent scattering
 on all plasma electrons and positrons can be expressed as
\begin{eqnarray}
S^{Pl} & = & \frac{ 16 i  c_e \alpha  m^2_e \pi^3}{\omega V {\it v
}_\phi }\, (\varepsilon^* \tilde F q) \, \delta^4(q - q') \,
\sum_{n=0}^\infty \, \int\limits_{-\infty}^{+\infty}
\frac{dp_3}{E_n} \,\{ f(E_n, \mu) + f(E_n, -\mu)\} \times
 \label{eq:Splasma} \\
& \times & \int\limits_0^\infty ds \left ( e^{is(q^2_\parallel +
2(pq)_\parallel)} + e^{is(q^2_\parallel - 2(pq)_\parallel)} \right
) \,
 e^{-iq^2_\perp sin(2\beta s)/2\beta} \,\, \lambda_n\left(\frac{q^2_\perp}{\beta} \, sin^2(\beta s)\right), \nonumber \\
& \qquad &  \lambda_n(x)  =   e^{-x} \{L_n(2x) - L_{n-1}(2x)\},
\nonumber
\end{eqnarray}
where $(pq)_\parallel = E\omega - p_3 q_3$, $L_n(x)$ is the Laguerre
polynomials, normalized by the  condition
 \bd \int\limits_{0}^{\infty} e^{-x} \, L^2_n(x) \, dx = 1. \ed

 As one can see from Eq.(\ref{eq:Splasma}), the four-dimensional
delta function, corresponding to the energy and momentum
conservation law is realized in the $S$-matrix element, as a
consequence of the fact that both initial and final states are the
neutral particles. Using the standard definition of an invariant
amplitude (\ref{eq:amp-def}) we can  write the plasma contribution
into the $\g$ in the  form:
\begin{eqnarray}
\g^{Pl} & = & \frac{ -2 i c_e \alpha m^2_e}{ \pi {\it v }_\phi }
\sum_{n=0}^\infty \,\, \int\limits_{-\infty}^{+\infty}
\frac{dp_3}{E_n} \,\{ f(E_n, \mu) + f(E_n, -\mu)\} \times
\label{eq:gplasma} \\
& \times & \int\limits_0^\infty ds \left ( e^{is(q^2_\parallel +
2(pq)_\parallel)} + e^{is(q^2_\parallel - 2(pq)_\parallel)} \right )
\,
 e^{-iq^2_\perp sin(2\beta s)/2\beta} \,\, \lambda_n \left(\frac{q^2_\perp}{\beta} \, sin^2(\beta s)\right). \nonumber
\end{eqnarray}

Notice that the result (\ref{eq:gplasma}) is valid for
electron-positron plasma in the presence of a constant magnetic
field of an arbitrary strength.

\section{The effective coupling between
pseudoscalar particle and photon in the strongly magnetized
plasma.}

\indent\indent In this section we investigate the familon-photon
interaction under the physical conditions when  from both
components of an active medium, a magnetic field  and plasma, the
magnetic field dominates, so the magnetic field strength appears
to be the largest physical parameter
\beq \beta \gg \mu^2, T^2, m^2_e. \label{eq:cond-1} \eeq

Such conditions could be realized, for example, in a supernova
explosion or in a coalescence of neutron stars where an outside
region of the neutrinosphere with strong magnetic field up to the
$10^{14}-10^{16}$ G and rather rarified plasma can exist.

The field contribution into effective coupling (\ref{eq:gfield2})
in the strong magnetic field limit ($\eta = B/B_e \gg 1$) can be
reduced to the form
\beq g^{F}_{\phi\gamma} =\frac{2 c_e \alpha}{\pi {\it v }_\phi }
\,\, H(z), \,\,\,H(z) = \int\limits_0^1 \frac{du}{1 - z(1 - u^2) }
- 1 , \label{eq:gfield-strong}\eeq
where $z= q^2_\parallel/4m^2_e$.

As for the plasma contribution to the effective coupling $\g$,
under the conditions (\ref{eq:cond-1}) when plasma electrons and
positrons occupy the ground Landau level only ($n=0$), we have
from  Eq.(\ref{eq:gplasma})
\begin{equation}
g^{Pl}_{\phi\gamma}  \simeq   \frac{- 4 c_e \alpha  m_e^2}{\pi
{\it v }_\phi} \,\,q_{\parallel}^2 \,
 \int\limits_{-\infty}^{+\infty} \frac{dp_3}{E} \,
  \frac{f(E,\mu) + f(E, -\mu)}{4(pq)_{\parallel}^2 - q_{\parallel}^4},
\label{eq:gplasma-strong}
\end{equation}
where  $E = \sqrt{p_3^2 + m^2}$  is the electron (positron) energy
on   the ground Landau level.

With  Eqs. (\ref{eq:gfield-strong}) and (\ref{eq:gplasma-strong}),
the effective coupling $\g$ in the strongly magnetized plasma is
\begin{eqnarray} g_{\phi\gamma} & = & \frac{ - 2 c_e \alpha}{\pi {\it v }_\phi} \,
F(q_\parallel), \label{eq:gstrong-tot}\\
F(q_\parallel) & =& 2m_e^2q_{\parallel}^2
\int\limits_{-\infty}^{+\infty} \frac{dp_3}{E} \,
  \frac{f(E,\mu) +  f(E,-\mu)}{4(pq)_{\parallel}^2 -
  q_{\parallel}^4}) - H(z).\nonumber
\end{eqnarray}

 It is interesting to note, that the
 expression (\ref{eq:gstrong-tot}) is valid also for the
familon (photon) propagating along the field direction
($q_\perp=0$) independently on the value of an external magnetic
field.

Really, considering $q_\perp=0$ in the field contribution into the
effective coupling  (\ref{eq:gfield2}) we immediately reproduce
the result (\ref{eq:gfield-strong}) obtained for the  ground
Landau level. As for  the plasma contribution described by
 the expression (\ref{eq:gplasma}), taking into account that
\beq \lambda_n(0) =
   \left \{ \begin{array}{ll} 1, & \, n = 0,
\nonumber\\
             0, &  n \neq 0,  \end{array} \right.
             \nonumber
             \eeq
one can see that the effective coupling
$\g$ is caused by the ground Landau level only.

So, the result  (\ref{eq:gstrong-tot}), obtained  for strongly
magnetized plasma when  plasma electrons and positrons occupy the
ground Landau level only has a more wide area of application. It
may be used even in the magnetic field of arbitrary  strength for
the familon (photon) propagating along the magnetic field.

 The expression (\ref{eq:gstrong-tot}) is significantly
simplified in some  limiting cases:
\begin{itemize}

\item the case of relatively high  familon (photon) energy
 ($\beta \gg q_\parallel^2 \gg m^2_e$). In this limit the plasma
 contribution  is suppressed by the electron mass $m_e$, which is the
 smallest
 parameter of the problem and
 the effective coupling $\g$ is caused mainly by the field
 contribution
\beq g_{\phi\gamma} \simeq g^{F}_{\phi\gamma} \simeq \frac{- 2 c_e
\alpha }{ \pi {\it v }_\phi}. \label{eq:g1}\eeq

\item  the case of relatively small familon (photon) energy  ($\omega^2 \ll
m_e^2$) when familon propagates transversely on the magnetic field
direction ($q_3=0$). In view of the asymptotic behavior of the
function $H(z)$ at small argument, the major contribution into the
effective coupling $\g$ comes  from plasma
\beq
 \g \simeq \g^{Pl} = \frac{ 2 c_e \alpha }
{\pi {\it v }_\phi}\int\limits_{m_e}^{\infty} dE \, \frac{p_3
\,}{E} \,
 \frac{d}{dE} \, (f(E,\mu) + f(E, -\mu)).
\label{eq:g2} \eeq

\item
the case of a strongly magnetized degenerate ultrarelativistic
plasma when familon energy $\omega < 2\mu$
\beq
 \g \simeq  \frac{-2 c_e \alpha}{\pi {\it v}_\phi}.
\label{eq:g3} \eeq
\end{itemize}

 As one can see from  Eqs.(\ref{eq:g1}), (\ref{eq:g2}) and (\ref{eq:g3}),  under some
 physical conditions  the effective  coupling $\g$ does not
 dependent on the  particle four-momentum.

\section{The conversion of the familon into two photons. }

\indent\indent The familon, being  a stable particle in the
vacuum, in the presence of a magnetized plasma can decay both into
the electron-positron pair in tree-level~\cite{Mikheev2} and into
the photon pair in the loop level. The decay $\phi \to \gamma +
\gamma$ becomes possible owing to the effective coupling $\g$ of
the familon-photon interaction  in a magnetized plasma. Really,
the Lagrangian of the familon-two photon  interaction can be
uniquely restored from (\ref{eq:lag1}) in the special case when
the effective coupling $\g$ is a constant independent on particle
momentum
 \beq L_{\phi\gamma\gamma} = \frac{\g}{4} \
(\tilde F^{\mu\nu} \,F_{\nu\mu}) \, \Phi, \label{eq:ad-lag} \eeq
where $F^{\alpha\beta}$ and $\tilde F^{\alpha\beta}=\frac{1}{2}
\varepsilon^{\alpha\beta\sigma\tau}F_{\sigma\tau}$ are  the tensor
and dual tensor of the quantized electromagnetic field.

In this section we investigate the conversion of a familon into
two photons in the strongly magnetized degenerate
ultrarelativistic plasma
\beq \beta \gg \mu^2 \gg T^2, m_e^2. \label{eq:cond_ph1}\eeq
assuming that the familon's energy  satisfies the condition
\beq  2\mu  > \omega \gg T. \label{eq:cond_ph2}\eeq

Notice, that under the condition (\ref{eq:cond_ph2}) the process
of the familon tree-level decay into electron-positron pair is
suppressed by the statistical factors in strongly degenerate
plasma.

In the strongly magnetized plasma as in the  strong  magnetic
field there are two photons modes
\beq \varepsilon_\alpha^{(1)} =
\frac{(q\varphi)_\alpha}{\sqrt{q^2_\perp}}, \qquad
\varepsilon_\alpha^{(2)} =
\frac{(q\tilde\varphi)_\alpha}{\sqrt{q^2_\parallel}}. \qquad \eeq
So, in a general case the following channels of the familon
conversion into the photon pair are possible: $\phi \to
\gamma^{(1)} + \gamma^{(1)}$, $\phi \to \gamma^{(1)} +
\gamma^{(2)}$, $\phi \to \gamma^{(2)} + \gamma^{(2)}$.

To obtain the decay probability one has to analyze the photon
dispersion in a magnetized plasma. By virtue of the fact that
photon of the first mode does not interact with electrons and positrons
occupying  the lowest Landau level, the dispersion law for the
first mode is the same as in a pure magnetic field~\cite{Shabad}
$$ q^{2} = \frac{-\alpha}{3 \pi} \, q^2_\perp.$$

As for the second photon mode, it turn out that its dispersion law
is defined by the same function $F(q_\parallel)$ as the
familon-photon effective coupling in the strongly magnetized
plasma (\ref{eq:gstrong-tot})
\beq q^{2} = \frac{2 \alpha eB}{ \pi} \, F(q_\parallel).
\label{eq:q}
\eeq

The function $F(q_\parallel)$ defined in Eq.(\ref{eq:gstrong-tot})
 tends to unity, $F(q_\parallel) \simeq 1$
 in the case of strongly magnetized degenerate ultrarelativistic plasma, so

$$ q^{2} = \frac{2 \alpha eB}{ \pi}.$$

Since the massless familon can decay only into photons  with $q^2
< 0$ under conditions considered, see Eqs. (\ref{eq:cond_ph1}) and
(\ref{eq:cond_ph2},) the familon conversion into two photons of
the first mode, $\phi \to \gamma^{(1)} + \gamma^{(1)}$,
 is the only possible channel of the familon decay.

The amplitude of this decay can be immediately obtained from the
Lagrangian (\ref{eq:lag-2}) and has the form
\beq
 M
 = \g \, \frac{(q_1 \varphi  q_2 )\,(q_1 \tilde \varphi q_2)}
{\sqrt{q^2_{1\perp} \, q^2_{2\perp}}} , \eeq
where $\g$ is described by the Eq.(\ref{eq:g3}), $q^\mu =
(\omega, \vec q)$, $q_1^\mu = (\omega_1, \vec q_1)$, $q_2^\mu =
(\omega_2, \vec q_2)$ are the familon and the final photons
4-momenta.

The probability  of the process $\phi \to \gamma^{(1)} +
\gamma^{(1)}$ under the conditions (\ref{eq:cond_ph2}) is defined
as
\beq \omega \,W = \frac{|M|^2}{64\, \pi^2} \,  \,
 \frac{d^3q_1}{\omega_1}\, \frac{d^3q_2}{\omega_2} \,
\delta^4(q - q_1 - q_2). \eeq

After integration over the phase space of the photons we find
\beq W  \simeq   \frac{\g^2  \alpha^2 \omega^3\, \sin^4 \theta
}{1152 \pi^3},
 \label{eq:ph_prob} \eeq
where $\theta$ is the angle   between the magnetic field direction
and the vector $\vec q$.

 Notice that the result (\ref{eq:ph_prob}) can be applied to
any pseudoscalar particle with the coupling  of the type
(\ref{eq:lag-2}). However, in the general case, it is necessary to
take into account the  pseudoscalar-photon coupling in vacuum due
to  the Adler's anomaly.

\section{Conclusion.}

\indent\indent We have studied the interaction between
pseudoscalar particle and photon in an electron-positron  plasma
with the presence of a constant magnetic field.  As an example of
a pseudoscalar particle, we have considered the familon,
associated with the spontaneous breakdown of a horizontal family
symmetry. The most general expression for the plasma and
 field contributions
into the effective coupling of familon-photon interaction in a
magnetized plasma was obtained. It was found that under some
physical conditions the effective coupling is a constant
independent on particle 4-momentum. In particular, the strong
magnetic field limit  was considered when plasma electrons and
positrons occupy the ground landau level only. It was shown that
the result for the effective familon-photon coupling obtained for
the strongly magnetized plasma is valid  in the case of arbitrary
magnetic field strength for the pseudoscalar particle propagating
along a magnetic field as well.

As an example of possible application of the result obtained we
have calculated the probability of the familon decay into two
photons under physical conditions of
strongly magnetized degenerate
ultrarelativistic plasma.

We believe that the result presented in this paper  would be
useful for the investigations of interaction between pseudoscalar
particle and photon in an active medium.

\vspace{10mm}

This work supported in part by the Council on Grants by the President of
 Russian Federation for the Support of Young Russian Scientists and Leading
 Scientific Schools of Russian Federation under the Grant No.
 NSh-1916.2003.2 andl
 by the Russian Foundation for Basic Research under the Grant No. 04-02-16253.

%\newpage

\end{document}